\documentclass[aps,prl,twocolumn,titlepage,nofootinbib]{revtex4}
\usepackage{graphicx}
\usepackage{bm}
\usepackage{epsfig}
\usepackage{ulem}
\usepackage{color}
\usepackage{mathrsfs}
\usepackage{dcolumn}
\usepackage{setspace}
\usepackage{array}
\usepackage{amsmath}
\usepackage{amssymb}
\usepackage{dsfont}
\usepackage{gensymb}
\usepackage{dcolumn}
\usepackage{multirow}
\usepackage{bibentry,natbib}
\usepackage{booktabs}

\begin{document}
\title{Angular Momentum-Dependent Spectral Shift in Chiral Vacuum Cavities}
\author{Qing-Dong Jiang$^{1,2,3}$}
\email{qingdong.jiang@sjtu.edu.cn}
\affiliation{{}\\ $^1$Tsung-Dao Lee Institute \& School of Physics and Astronomy, Shanghai Jiao Tong University, Shanghai 200240, China\\
$^2$Shanghai Branch, Hefei National Laboratory, Shanghai 201315, China\\
$^3$Shanghai Research Center for Quantum Sciences, Shanghai 201315, China
}
\begin{abstract}
Based on a hybrid light-matter unitary transformation for cavity quantum electrodynamics, we investigate the spectral shift of an atom induced by quantum fluctuations in a chiral vacuum cavity. Remarkably, we find an intriguing angular momentum-dependent shift in the spectra of bound states. {Our approach shows promise in going beyond traditional perturbative methods and demonstrates effectiveness even in the strong-coupling limit, as evidenced by our numerical benchmarks in the case of a two-dimensional quantum harmonic oscillator.} In addition, we establish a cavity-interaction picture for calculating the chiral vacuum Rabi oscillation in the strong-coupling limit for a generic central potential. The anomalous spectral shift revealed in this study possesses both fundamental and practical significance and could be readily observed in experiments.
\end{abstract}
\maketitle


\section{I. Introduction}
Vacuum is not void; instead, it is full of quantum fluctuations with {virtual} particles constantly being created and annihilated. The vacuum quantum fluctuations give rise to a plethora of well-known phenomena, including the Casimir effect \cite{casimir1948attraction,bordag2001new,milton2004casimir,plunien1986casimir}, the Lamb shift \cite{lamb1947fine,bethe1947electromagnetic,maclay2020history}, anomalous magnetic moment \cite{PhysRev.82.664,weinberg1995quantum}, vacuum Rabi oscillations \cite{jaynes1963comparison,vacuumrabi}, and spontaneous photon emission \cite{dalibard1982vacuumradiation}. 
In addition to these fundamental effects, physicists have directly probed the electromagnetic fluctuations within a vacuum cavity (often referred to as a ``dark cavity") \cite{riek2015direct,benea2019electric}. The cavity offers a notable advantage as it allows for significant amplification of the quantum fluctuations by squeezing the cavity volume \cite{PhysRevA.43.398,ikuta2021cavity,garziano2015multiphoton,shapiro2015dynamical}.

In recent years, researchers have achieved remarkable success in creating extremely small cavities, approaching the nanoscale \cite{nanocavisong,epstein2020far,bylinkin2021real}. These advancements have paved the way for exploring the realm of strong light-matter coupling across various setups. 
Comparing to the Floquet method (i.e., engineering material properties with electromagnetic radiations), using cavity quantum fluctuations for material property engineering has obvious advantages \cite{jiang2023engineering,schlawin2022cavity,hubener2021engineering,bloch2022strongly,PhysRevApplied.18.044011,schafer2018ab,bacciconi2023first,rokaj2023cavity,mercurio2023photon}:
i) Within the cavity, the interactions between light and matter surpass the limitations imposed by classical light-matter interaction bounded by the fine structure constant. As a result, the properties of materials can be deeply tailored in cavities { \cite{paravicini2019magneto}}. 
ii) Engineering materials and molecules within vacuum cavities is superior to the { Floquet engineering} where external electromagnetic radiation heat up the system and destroy quantum effects { \cite{jarc2023cavity}}.
iii) Cavity quantum fluctuations allow for the engineering of material properties in an equilibrium manner, in contrast to the {  Floquet engineering} that drives the system out of equilibrium, resulting in transient and complex physical properties {\cite{appugliese2022breakdown}}. 

Over the past several years, researchers have presented pioneering proposals, with some already realized, to utilize cavity quantum fluctuations for engineering material conductivity \cite{rokaj2022free,moddel2021casimir,cardoso2025cavity}, inducing anomalous superconductivity \cite{sentef2018cavity,schlawin2019cavity,curtis2019cavity,thomas2019exploring}, changing band structure, topology and magnetism\cite{PhysRevB.104.155307,appugliese2022breakdown,PhysRevB.105.165121,rokaj2023topological,jiang2023engineering,yang2024emergent,wei2024cavity}, and even modulating chemical reactivity \cite{PhysRevLett.116.238301,flick2017atoms,galego2017many,galego2019cavity,schafer2019modification,altman2021quantum,PhysRevB.103.165412}. These advancements exemplify the profound potential of cavity quantum fluctuations in tailoring material properties.

Nevertheless, despite its advantages, using cavity quantum fluctuations to control quantum states of matter faces a major obstacle. Unlike real electric or magnetic fields, most quantum fluctuations inherently maintain parity symmetry (PS) and time-reversal symmetry (TRS). As a result, their ability to manipulate material and molecular properties is constrained. To induce substantial changes in material properties, it becomes essential to encode symmetry breaking into quantum fluctuations.
In recent years, multiple works have shown the impact of discrete symmetry breaking on phenomena induced by quantum fluctuations. Notable examples include symmetry breaking induced anomalous Casimir forces \cite{butcher2012casimir,jiang2019axial,dai2024universal}, {chirality selection} in chemical reactions \cite{ke2023vacuum,riso2022strong}, and topological response and phase transitions \cite{espinosa2014semiconductor,PhysRevB.99.235156,PhysRevB.106.205114,jiang2023engineering,yang2024emergent,yang2025quantum}. 
A recent work by Wilczek and the author \cite{jiang2019quantum} highlighted the combined power of symmetry breaking and quantum fluctuations. It shows that symmetry breaking can be transmitted from materials to their vicinity by vacuum quantum fluctuations. The vacuum in proximity to a symmetry-broken material was referred to as its {\it Quantum Atmosphere}. {For instance, a material exhibiting spontaneous time-reversal symmetry breaking will impart this property to the nearby vacuum through quantum fluctuations. Consequently, an atom situated within a cavity is enclosed by time-reversal symmetry breaking quantum fluctuations, as illustrated in Figure 1."}
\begin{figure}[!htb]
\includegraphics[height=6.6cm, width=7.2cm, angle=0]{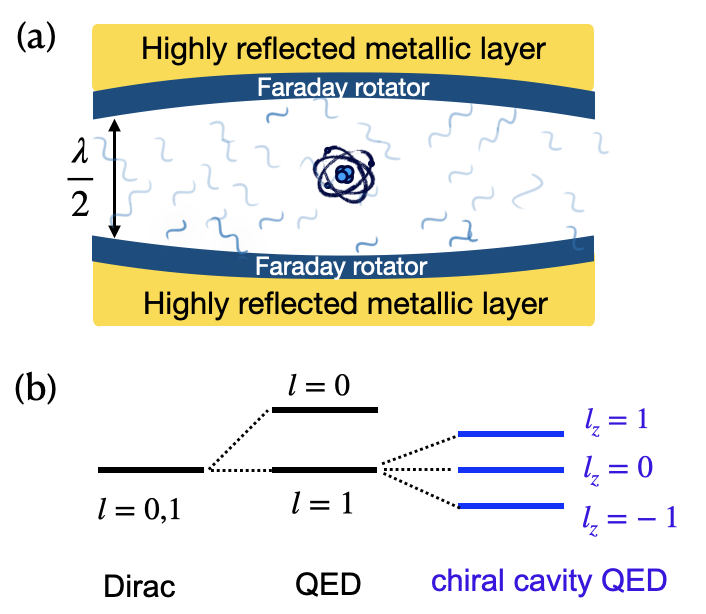}
 \caption{{(a) The schematic setup of an atom embedded in a chiral cavity, created by depositing a layer of Faraday rotator material on a metallic substrate. (b) Illustration of the spectral shift induced by chiral cavity QED.
 }}\label{fig1}
\end{figure}

In this paper, we present what, to the best of our knowledge, are the first fully quantum mechanical predictions of angular momentum(AM)-dependent spectral shifts, induced by the quantum fluctuations in a chiral cavity. 
{ While (chiral) cavity QED has been previously explored both experimentally and theoretically \cite{zhang2019chiral,mivehvar2021cavity,leonard2017supersolid,landig2016quantum,sauerwein2023engineering}, calculations of the Lamb shift are still lacking for both strong light-matter coupling and time-reversal symmetry breaking cases. In this study, we successfully derive rigorous analytical results for both situations.}
{ Additionally, we employ numerical exact diagonalization to verify the effectiveness of our analytical predictions in the intermediate coupling regime.}
{Chiral cavity break time-reversal symmetry, favoring only one type of handedness of photons. A straightforward method for achieving an effective chiral cavity involves utilizing a Faraday rotator (\textit{e.g.}, ferromagnetic layer) in conjunction with high-quality metallic mirrors to establish a high quality cavity  \cite{voronin2022single,baranov2023toward,hubener2021engineering}.Our major finding is that we are able to calculate the spectral shift of an atom induced by the quantum fluctuations in a chiral cavity in the strong coupling limit. 
It is also worth noting that our analytical calculation of the cavity-Lamb (CL) shift in the strong coupling limit should be equally valuable.} In the last part, we establish a framework---the cavity-interaction picture---to calculate time-dependent phenomena. This enables us to compute chiral vacuum Rabi oscillation in the strong coupling limit for a generic central potential. 
{ For simplicity, we assume a cavity with high quality factor and ignore the influence of cavity losses. An in-depth analysis of how these factors would influence the outcomes is detailed in the subsequent sections of this manuscript and the Supplemental Materials (SM) \cite{supplementalMat}.}

\section{II. Chiral unitary transformation and cavity-induced potential shift}

To set the stage, we examine the generic Hamiltonian 
\begin{equation}\label{hamilton1}
\hat H= \frac{1}{2m}\left(\hat{\bold p}-q\hat{\bold A}\right)^2+V(\bold r)+\hbar \omega_c\hat a^\dagger \hat a,
\end{equation}
which captures the interaction between a charged particle (with mass $m$, charge $q$) and a photonic mode in a cavity. 
We assume an external single-particle potential, $V(\bold r)$, and a single photonic mode with frequency $\omega_c$, where $\hat a$ and $\hat a^\dagger$ are the annihilation and creation operators of photons, respectively. 
{
The following discussion can be generalized to multi-mode cases. We present the results in the discussion section and provide detailed calculations in the SM \cite{supplementalMat}.}
The vector potential $\hat{\bold A}$ can be expressed as $\hat{\bold A}=A_0\left(\boldsymbol \varepsilon^* \hat a^\dagger+\boldsymbol\varepsilon \hat a\right)$, where $\boldsymbol \varepsilon$ represents the polarization of the cavity photonic modes. The mode amplitude is  $A_0=\sqrt{\frac{\hbar}{2\epsilon_0 V\omega_c}}$, with $V$ the cavity volume. 
In the context of cavity quantum electrodynamics, a dimensionless parameter $g=\sqrt{\frac{(q A_0)^2}{m\hbar \omega_c}}$ is commonly used to quantify the strength of the light-matter coupling. The regime $10^{-1} \leqslant g \leqslant 1$ is referred to as strong coupling. For $g \geqslant 1$, it is referred to as deep strong coupling, indicating even stronger interaction between light and matter\footnote{The terminology differs slightly from the standard context of quantum optics, where the term ``strong coupling" typically refers to reversible interactions between photons in the cavity mode and the atom. Here, ``strong coupling" and ``weak coupling" signify the relative strength of the light-matter coupling in comparison to the cavity mode energy.}.
In this letter, we focus on the chiral cavity case, where the photonic polarization is $\boldsymbol\varepsilon=\frac{1}{\sqrt{2}}\left(\mathbf{e_x}+i\mathbf{e_y}\right)$, and $\mathbf{e_{x(y)}}$ represents the unit vector in the x(y)-direction. 
A recent seminal advancement in cavity quantum electrodynamics is the ability to decouple matter and light degrees of freedom, either in the weak or strong coupling limits, through a special unitary transformation \cite{ashida2021cavity,PhysRevB.107.195104}. This transformation is elegantly achieved by applying the unitary operator:
\begin{equation}
\hat U=\exp\left[-i \frac{\xi}{\hbar} \hat{\bold p}\cdot \hat{\boldsymbol\pi}\right],~{\text{with}}~\xi=\frac{g}{1+g^2}\sqrt{\frac{\hbar}{m\omega_c}}
\end{equation}
to the original {Hamiltonian $H$}, where $\hat{\boldsymbol \pi}=i\left(\boldsymbol \varepsilon^* \hat a^\dagger-\boldsymbol \varepsilon \hat a\right)$ is the photonic momentum operator. The parameter $\xi$ is chosen to eliminate the linear light-matter coupling term ($\hat{\bold p}\cdot \hat{\bold A}$). Remarkably, this unitary transformation yields an equivalent yet formally much neater Hamiltonian:
\begin{eqnarray}\label{hamiltonU}
\hat H^\prime(\xi)&=&\hat U^\dagger \hat H \hat U\nonumber\\
&=&\frac{\hat{\bold p}^2}{2m_{\text{eff}}}+V\left(\bold r+\xi \hat{\boldsymbol \pi}+\frac{\xi^2}{2\hbar }\hat{\bold p}\times \bold{e_z}\right)+\hbar \omega_{\text{eff}}\hat a^\dagger\hat a\nonumber\\
\end{eqnarray}
with the renormalized mass $m_{\text{eff}}=m (1+g^2)$ and the effective cavity frequency $\omega_{\text{eff}}=\omega_c(1+g^2)$.
It is important to note that the light-matter coupling is fully encapsulated in the shifted single-particle potential, offering a key advantage of the transformed Hamiltonian (Eq.\eqref{hamiltonU}).
Several remarks are in order to better understand the above transformation: 
\begin{itemize}
\item The key advantage of Eq.\eqref{hamiltonU} is that the light-matter coupling is fully encoded in the shifted single-particle potential. 

\item
The coupling parameter $\xi=\frac{g}{1+g^2}\sqrt{\frac{\hbar}{m\omega_c}}$ approaches zero not only in the weak-coupling limit ($g\rightarrow 0$) but also in the strong-coupling limit ($g\rightarrow \infty$). 

\item
Cavity light-matter interactions lead to an increase in both the effective mass of the particle ($m_{\text{eff}}> m$) and the effective mode frequency ($\omega_{\text{eff}} >\omega_c$). 
\end{itemize}
These features allow the application of perturbation theory (in terms of $\xi$) to investigate strong light-matter coupling in cavities. 
{ It is worth noting that the asymptotic decoupling of light and matter degrees of freedom may be fundamentally connected to the disentanglement of light and matter in the strong-coupling limit, as discussed in the insightful review \cite{frisk2019ultrastrong}.} 
In the following sections, we explore several prominent effects induced by chiral cavities, including angular momentum-dependent spectral shift, the cavity Lamb shift, and chiral vacuum Rabi oscillations.

\section{III. Cavity QED renormalized spectra} 

We now examine the influence of quantum fluctuations in a cavity on the spectral shift of a bound state governed by the Hamiltonian Eq.\eqref{hamilton1}. 
{ For simplicity, let us consider a central potential $V(\bold r)=V(r)$. 
The potential term in Eq.(3) can be written as
$\hat V\left(\bold r+\boldsymbol{\hat \tau_c}\right)=\hat V\left(r\right)+\Delta{\hat V}$,
where $\boldsymbol{\hat{\tau_c}} = \xi \hat{\boldsymbol{\pi}} + \frac{\xi^2}{2\hbar} \hat{\bold{p}} \times \bold{e_z}$ and $\Delta \hat{V}$
can be expanded in terms of $\tau_c$ (assuming small $\xi$):
\begin{eqnarray}
\Delta{\hat V}
\approx \boldsymbol{\hat \tau_c}\cdot \boldsymbol\nabla V(r)
+\frac{1}{2}\left(\boldsymbol{\hat \tau_c}\cdot \boldsymbol\nabla\right)^2V(r).
\end{eqnarray}}
Consequently, the transformed Hamiltonian reads
\begin{eqnarray}
\hat H^\prime=\frac{\hat{\bold p}^2}{2m_{\text{eff}}}+\hat V\left(r\right)+\hbar \omega_{\text{eff}}\,\hat a^\dagger\hat a+\Delta {\hat V},
\end{eqnarray}
which applies to both weak and strong light-matter interactions. 
{
In the intermediate regime where the perturbative expansion is inapplicable, Eq. (5) may offer a distinct advantage for computing the renormalized spectrum through exact diagonalization in cavity QED (see, for example, \cite{hubener2021engineering} and references therein).} However, this task remains to be done in the future.
With the above preparation, we can employ perturbation theory using the unperturbed states, which are the product states of the n-th bound state $|\psi_n\rangle$ and the cavity vacuum state (zero photon) $|0\rangle_{\text{cav}}$, i.e.,
$|\Psi_{n}\rangle=|\psi_n\rangle\otimes |0\rangle_{\text{cav}}$, where $|\psi_n\rangle$ represents the bound state of a particle with an effective mass $m_{\text{eff}}$ in a central potential $V(r)$. 
{
According to the definition of effective mass in Eq. (3), the effective mass approaches the bare mass in the weak-coupling limit, while it significantly deviates from the bare mass in the strong light-matter coupling regime.
}
The first-order perturbation calculation yields the energy shift:
\begin{eqnarray}
\Delta E_{n}=\langle \Psi_{n}|\Delta \hat V |\Psi_{n}\rangle=\Delta E_{n}^{\text{AM}}+\Delta E_{n}^{\text{CL}},
\end{eqnarray}
where $\Delta E_n^{\text{AM}}$ and $\Delta E_n^{\text{CL}}$ are the angular momentum (AM)-dependent shift and the cavity-Lamb (CL) shift, respectively. They are given by
\begin{eqnarray}
\label{ameq}\Delta E_n^{\text{AM}}&=&\frac{\xi^2}{2\hbar}\langle \psi_n|\frac{1}{r}\frac{dV(r)}{dr}\hat L_z|\psi_n \rangle\\
\label{cleq}\Delta E_n^{\text{CL}}&=&\frac{\xi^2}{4}\langle \psi_n |\nabla^2 V(r)|\psi_n \rangle.
\end{eqnarray}
The expressions, Eq.\,(8) and Eq.\,(9), are the key findings of this letter and remain applicable in both the weak and strong coupling regimes.
Eq.(8) indicates that the cavity quantum fluctuations indeed encode the breaking of time-reversal symmetry. 
This is because, in the presence of time-reversal symmetry, states with opposite angular momentum, $l_z=\pm 1$, would have the same energy. 
Notably, reversing the cavity's chirality induces a sign change in the AM-dependent spectral shift, showing the essential importance of the cavity's chirality.
In what follows, we will provide examples to illustrate these two key formulas and demonstrate they predict directly measurable effects.

{\section{IV. Anomalous spectral shift in two examples}}---
We evaluate the AM-dependent shift and the CL shift in two examples: the Hydrogen atom and the two-dimensional harmonic oscillator. Let us first focus on the spectral shift of the Hydrogen atom model to gain physical understanding. 
For the Hydrogen atom, with the potential $V(r)=-k/r$ (where $k\equiv e^2/4\pi \epsilon_0$), we determine the spectral shifts for each energy level. 
By substituting the eigen function of the Hydrogen atom into the formulas, we obtain the spectral shifts of the bound state $|\psi_{n,l,l_z} \rangle$, where $n$, $l$, and $l_z$ are the principal, azimuthal, and magnetic quantum numbers, respectively:
\begin{eqnarray}
\Delta E_{n,l,l_z}^{\text{AM}}&=&
\frac{l_z\,\xi^2 \,k}{2 a_{\rm eff}^3 n^3 l(l+\frac{1}{2})(l+1)}; \\
\Delta E_{n,l,l_z}^{\text{CL}}&=&
\frac{\pi \xi^2\, k}{n^3 a_{\rm eff}^3}\,\delta_{l,0}\,\delta_{l_z,0}.
\end{eqnarray}
where $a_{\rm eff}=4\pi\epsilon_0 \hbar^2/m_{\rm eff}e^2$ is the effective Bohr radius. In these calculations, we have used the relations $\langle 1/r^3\rangle=1/a_0^3 n^3 l(l+1/2)(l+1)$ and $\nabla^2 V=4\pi k \delta(r)$.
The spectral shifts can be easily estimated. For instance, the AM-dependent spectral shift of the first excited state with angular momentum $l=1$ and $l_z=\pm 1$ is given by $\Delta E_{2,1,\pm 1}= \pm \left(\frac{\xi}{a_{\text{eff}}}\right)^2 \frac{{\text{Ry}}}{24}\frac{m_{\text{eff}}}{m}\approx 0.3\, \text{meV}$, where $g=0.01$, $\omega_c=10^{16} s^{-1}$, and $\rm Ry$ is the Rydberg energy. This estimation assumes a single-mode scenario, but it can be extended to include multiple modes (see later). Furthermore, we can recover the Lamb shift by considering a large cavity (i.e., weak light-matter coupling limit) and integrating over all possible mode frequencies.
It yields
\begin{eqnarray}
\Delta E^{\rm Lamb}&=&\sum_n\frac{\hbar}{m\omega_{c,n}}\frac{g^2}{2} \langle \Psi_n|\nabla^2 V(r) |\Psi_n\rangle\nonumber\\
&=&\frac{1}{8\epsilon_0\pi^2}\int d\omega_c   \frac{\hbar}{\omega_{c}}\frac{q^2}{m^2} \langle \Psi_n|\nabla^2 V(r) |\Psi_n\rangle\nonumber\\
&=&\frac{\hbar q^2}{8\epsilon_0\pi^2 m_{\rm eff}^2}\ln{\frac{1}{\pi \alpha}} \langle \Psi_n|\nabla^2 V(r) |\Psi_n\rangle.
\end{eqnarray}
Here, in line with Hans Bethe's approach \cite{bethe1947electromagnetic}, we have regularized the non-relativistic theory by selecting $\hbar \omega_{\text{min}}=\hbar c \pi/a_0$ (where $a_0$ represents the Bohr radius) as the smallest energy scale and $\hbar \omega_{\text{max}}=mc^2$ as the largest energy scale. We remark that the derivation of the Lamb shift closely resembles Theodore A. Welton's approach in the weak-coupling limit \cite{welton1948some}. 
{ Note that when expanding the modified potential in terms of $\xi$, an additional contribution proportional to  $(\xi \vec{\pi}\cdot \nabla V)^2$ also arises, contributing to the traditional Lamb shift. However, since this term is proportional to $V^2$ and thus scales as $\alpha^2$, it represents a higher-order correction to the Lamb shift.}

Next, we consider the spectral shift of a two-dimensional (2D) harmonic oscillators governed by the Hamiltonian $\hat H=\frac{p_x^2+p_y^2}{2m}+\frac{m}{2}\omega^2\left(x^2+y^2\right)$. This Hamiltonian exhibits rotational symmetry and commutes with the angular momentum operator along the z-axis, $\hat L_z$. By introducing the annihilation operator 
$\hat a_{R(L)}=\left[\sqrt{\frac{m\omega}{\hbar}} (x\pm i y)+i\frac{p_x\pm p_y}{\sqrt{m\hbar\omega}}\right]/2$,
one can rewrite the Hamiltonian and angular momentum operator in terms of number operators $\hat n_{L(R)}=\hat a_{R(L)}^\dagger \hat a_{R(L)}$:
\begin{eqnarray}
    \hat H_{\rm HO}=\left(\hat n_R +\hat n_L  +1\right)\hbar \omega;~
    \hat L_z=\hbar \left(\hat n_R  -\hat n_L  \right).
\end{eqnarray}
Here $\hat H_{\rm HO}$ and $\hat L_z$ share the common set of eigenstates
\begin{equation}
    |\phi_{n_R,n_L}\rangle=\frac{1}{\sqrt{n_R!n_L!}}(a_R^\dagger)^{n_R}(a_L^\dagger)^{n_L}|\phi_{0,0}\rangle,
\end{equation}
where $n_R$ and $n_L$ are integers that characterize an eigenstate.
According to the Eq.(8), the AM-dependent spectral shift of the state $|\phi_{n_R,n_L}\rangle$ is given by
\begin{eqnarray}
\Delta E^{\text{AM}}=\frac{\xi^2}{2}m \omega^2\left(n_R-n_L\right). 
\end{eqnarray}
For the ground state of the 2D quantum Harmonic oscillator, $\langle \hat L_z\rangle_n=0$, and AM-dependent spectral shift vanishes. However, a spectral gap of size $m\omega^2\xi^2/2 $ emerges for the originally degenerated first excited states (i.e., $|\phi_{1,0}\rangle$ and $|\phi_{0,1}\rangle$) with different angular momentum. The CL shift remains a constant due to $\nabla^2 V(r)=m\omega^2$ in this special case. 
In addition to the two prototypical examples, our approach is applicable to a wide range of real experimental systems \cite{khitrova2006vacuum,toida2013vacuum}. For instance, one could measure the spectral shift of Rydberg atoms, superconducting circuits, quantum dots or excitons in transition-metal dichalcogenides, which can be effectively described by an hydrogen atom model \cite{PhysRevA.56.1443,gramich2014lamb,zhou2015berry,PhysRevLett.115.166802}. 
{ To assess the validity of our analytical results in the strong coupling regime (i.e., $g \sim 1$), we perform exact numerical diagonalization in an extended Hilbert space that includes both electronic and photonic degrees of freedom. Remarkably, the numerical calculations serve as a direct benchmark, confirming the effectiveness of our analytical predictions, even in the strong coupling regime (see numerical benchmarks section). }

\section{V. Cavity interaction picture and polaritonic vacuum oscillation}

Spectral shifts and spontaneous emission are interconnected consequences of quantum fluctuations. In a vacuum cavity, an excited atom can spontaneously emit and reabsorb a cavity photon, a phenomenon known as vacuum Rabi oscillation \cite{fox2006quantum,PhysRevLett.76.1800,meunier2005rabi}.
In this section, we investigate vacuum Rabi oscillation in a chiral cavity, examining both the weak and strong light-matter coupling regimes. To proceed, we introduce the {\it cavity interaction picture}: In the cavity-interaction picture, quantum states and operators are defined as follows:
$|\Psi(t)\rangle_I=e^{{i\hat H_0 t}/{\hbar}}|\Psi(t)\rangle$ and $\Delta \hat V_{\mathrm{I}}(t)=e^{i \hat H_0 t} \Delta \hat V(t) e^{-i \hat H_0 t}$, where $|\Psi(t)\rangle$ and $\Delta \hat V(t)$ represent the quantum state and operator in Schrodinger picture.
{ In contrast to the traditional interaction picture, the Hamiltonian $H_0=\hat p^2/{2m_{\rm eff}}+V(r)$ in cavity-interaction picture has an effective mass $m_{\rm eff}=m(1+g^2)$ and includes potential $V(r)$.}
The interaction picture is highly useful for studying time-dependent phenomena. The wave function in the cavity interaction picture evolves according to $|\Psi(t)\rangle_{I}=\hat U_I(t,0)|\Psi(0)\rangle_{I}$, where the unitary evolution operator is given by
\begin{eqnarray}
\hat U_I(t,0)=\mathcal T\left\{\exp \left[-\frac{i}{\hbar}\int_0^t d\tau \Delta \hat V_I(\tau)\right]\right\}
\end{eqnarray}
where $\hat U_I(t,0)$ represents the time evolution operator and $\mathcal T$ stands for time ordering operator.

Based on the cavity interaction picture, let us consider a two-level system consisting of an excited state $|e\rangle$ and a ground state $|g\rangle$ within a vacuum cavity. Specifically, we focus on the lowest two levels of the combined system, which are represented by the product states $|\Psi_1 \rangle =|e\rangle |0\rangle_{\rm cav} $ and $|\Psi_2 \rangle =|g\rangle |1\rangle_{\rm cav} $, where $|0\rangle_{\rm cav}$ and $|1\rangle_{\rm cav}$ correspond to the cavity photon states with zero and one photon, respectively. The scattering matrix is then given by
\begin{eqnarray}
\Delta \hat V_{\mathrm{I}}(t)=\left(\begin{array}{cc}
\gamma_{11} & \gamma_{12}e^{-i \tilde \omega t}  \\
\gamma_{21}e^{i \tilde\omega t} &\gamma_{22}
\end{array}\right)
\end{eqnarray}
where $\gamma_{ij}=\langle \Psi_i| \Delta \hat V|\Psi_j\rangle$ and $\tilde\omega=\omega_2-\omega_1$ represent the spectral gap between the two states.
In the first-order approximation of the scattering matrices, the unitary evolution operator is given by
\begin{eqnarray}
\hat U_{I}(t, 0)=1-\frac{i}{\hbar}  \left(\begin{array}{cc}
\gamma_{11}t & \gamma_{12} \frac{\sin \left(\tilde{\omega} t\right)}{\tilde{\omega}} e^{-i \tilde{\omega} t} \\
\gamma_{21} \frac{\sin \left(\tilde{\omega} t\right)}{\tilde{\omega}}e^{i \tilde{\omega} t} & \gamma_{22}t
\end{array}\right)+\dots \nonumber
\end{eqnarray}
If the system is initially prepared in the first excited state, i.e.,
$|\Psi(0)\rangle_{I}=\left(0,~1\right)^T$,
then the wave function at later times is given by $|\Psi(t)\rangle_{I}=\hat U_I(t, 0) |\Psi(0)\rangle_{I}$.
Therefore, the probability of finding the system in the ground state after time $t$ is
\begin{equation}
    P_{1\rightarrow 2}\left(t, \tilde{\omega}\right) \equiv\left|\left\langle \Psi_2| \Psi_I(t)\right\rangle\right|^2=\gamma_{12}^2 \frac{\sin ^2\left(\tilde{\omega} t\right)}{\hbar^2\tilde{\omega}^2}
\end{equation}
where the scattering matrix is represented by
\begin{eqnarray}\label{scatmatr}
\gamma_{12}^{\rm AM}&=&-i\frac{\xi}{\sqrt{2}} \langle e |\left(\partial_x+i\partial_y\right)V(r)|g\rangle\nonumber\\
&=&-i\frac{\xi}{\sqrt{2}} \langle e |\frac{d V(r)}{d r}  e^{i\theta}|g\rangle. 
\end{eqnarray}
{
Note that in the above derivations, we have assumed that $\gamma_{12}\ll \tilde\omega$ to ensure no resonance transition occurs. Note that we have derived a general result that is applicable to both the weak and strong light-matter coupling regimes.} Eq.~\eqref{scatmatr} shows that the scattering matrix connects quantum states with magnetic quantum numbers that differ by exactly one $\hbar$.  This finding indicates that a chiral photon is emitted and re-absorbed in a chiral vacuum cavity.
 
\section{VI. Multi-mode effect}

{ The calculation for the single mode case can be generalized to the $N$-mode scenario, where the quantized vector potential is expressed as $\hat{\bold A}=\sum_k^N A_0\left(\boldsymbol\varepsilon \hat a_k +\boldsymbol \varepsilon^* \hat a_k^\dagger\right)$. To asymptotically decouple the light and matter degrees of freedom, we propose using a generalized unitary transformation:
\begin{equation}
\hat U=\Pi_k^N \exp\left[-i \frac{\xi_k}{\hbar} \hat{\bold p}\cdot \hat{\boldsymbol\pi_k}\right],
\end{equation}
where again $\hat{\boldsymbol \pi}_k=i\left(\boldsymbol \varepsilon^* \hat a_k^\dagger-\boldsymbol \varepsilon \hat a_k\right)$, and $\xi_k$ should be recalculated for the multi-mode case. Remarkably, we find the required $\xi$ for $N\geq 2$ multi-mode case is \cite{supplementalMat}:
\begin{eqnarray}\label{multieqtran}
\xi_i=\frac{\prod^N_{k\neq i}\omega_k\left(\frac{qA_0}{m}\right)}{\prod_{k=1}^N\omega_k+\left(\frac{q^2A_0^2}{2m}\right)\left[\sum_{j=1}^N \prod_{k\neq j}\omega_k\right]}. 
\end{eqnarray}
With this expression, the corresponding calculations can proceed using the modified $\xi$ for the multi-mode case.

}

{
\section{VII. Numerical Benchmarks}

Before concluding, we present the numerical benchmarks for our analytical calculations near $g \sim 1$ for both the hydrogen atom (Fig.2 (a)) and 2D harmonic oscillator (Fig. 2(b)) cases. We find that the numerical and analytical results align more closely for the 2D harmonic oscillator. This is likely due to the fact that, in the 3D hydrogen atom model, the cavity polarization lies in a 2D plane, which modifies the effective mass only in the x and y directions. For comparison with the exact numerical results, we used an averaged effective mass in our analytical hydrogen model. 
\begin{figure}[!htb]
\includegraphics[height=3.2cm, width=7.2cm, angle=0]{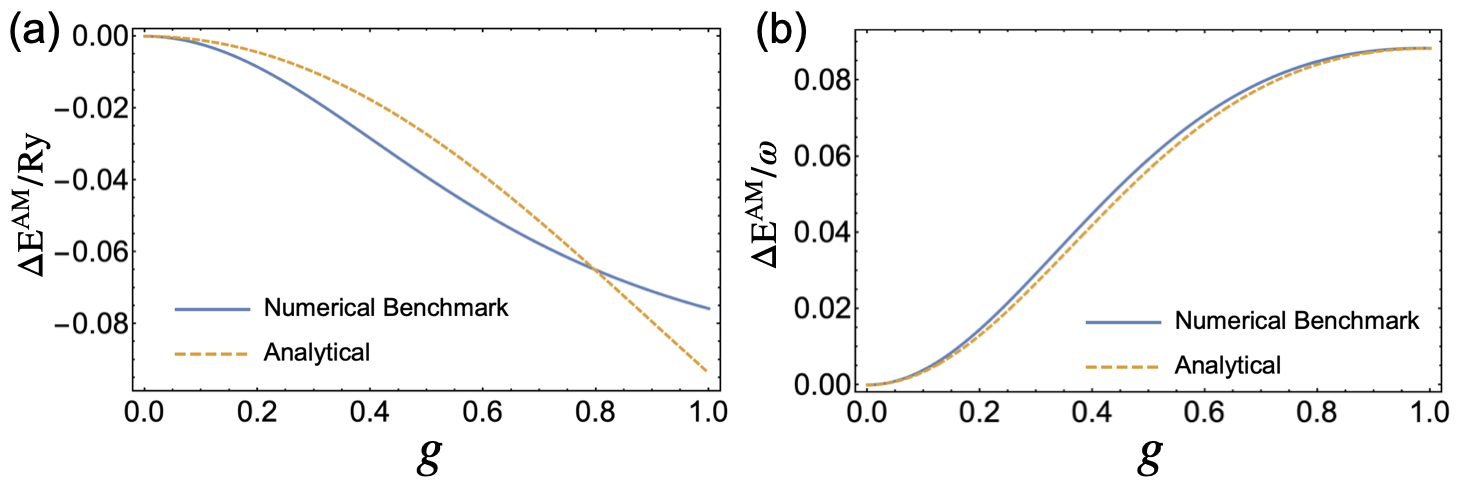}
 \caption{{ In (a) we compare the numerically calculated angular-momentum dependent spectral shifts for the hydrogen atom with the analytical result where the averaged effective mass $\tilde m_{\rm eff}=(2 m (1+g^2)+m)/3$ is used in our analytical formula. In (b) we show compare the numerical results and analytical results for 2D harmonic oscillator. The energy units are chosen to be the characteristic energy scales for each system, namely the Rydberg energy for the hydrogen atom and the frequency of the harmonic oscillator for the 2D case. 
 To ensure convergence of the numerical calculations, we diagonalize the Hamiltonian within a 
$1024\times 1024$-dimensional Hilbert space, constructed from the low-energy product states of the electronic and photonic Hamiltonians.
 }}\label{fig2}
\end{figure}
In contrast, the results for the 2D harmonic oscillator show a remarkable consistency between the analytical and numerical calculations.
 
}

\section{VIII. Summary}

In our analysis, we focused on examining the spectrum of a single atom. Nevertheless, it is worth noting that our approach works for many-body systems, wherein collective enhancement can be anticipated {\cite{rokaj2023cavity}}. For example, for a group of electrons subjected to a confining potential $V(r)$, the spectral shift scales with the total angular momentum of all electrons, i.e., $\sum_i\langle \hat L_z^{(i)} \frac{1}{r}\frac{d V}{dr}\rangle$, given that all electrons coherently couple to a single cavity mode.
Additionally, we should address the valid range of our perturbation theory. To apply our theory to cases involving strong light-matter coupling, it is necessary for the shift parameter $\xi$ to be considerably smaller than the typical length scale within the system. { For example, when applying our theory to atomic spectra, we require $\xi/a_{\rm eff}\ll 1$, where $a_{\rm eff}$ is the aforementioned effective Bohr radius of an electron with an effective mass, $m_{\rm eff}$. While for an electron in a vacuum, meeting this requirement becomes challenging
for strong cavity light-matter coupling due to $\xi/a_{\rm eff}=g\alpha \sqrt{\frac{mc^2}{\hbar \omega_c}}$, electrons in small-band semiconductors (e.g. InSb) can have a very small mass ($\sim 0.01 m_e$) and should easily satisfy this condition for $g\sim 0.1$. In contrast, for the 2D harmonic oscillator, the characteristic length scale remains unchanged, making it easier to satisfy the perturbation condition.}

In summary, we have successfully developed a perturbation theory applicable to both weak and strong light-matter coupling regimes, uncovering an AM-dependent chiral spectral shift in chiral cavities. We determined the AM-dependent spectral shift and CL shift for two specific examples, demonstrating that the effect is robust and detectable in experimental settings. Furthermore, we have established the foundation for cavity time-dependent perturbation theory, enabling us to calculate chiral vacuum Rabi oscillations for arbitrary central potentials in the regime of strong light-matter coupling.

{\textit{Acknowledgement}.---}
We appreciate the indispensable discussions and help from Liu Yang, Hans Hansson, Jianhui Zhou and Yi-Zhuang You. This
work was supported by the Innovation
Program for Quantum Science and Technology Grant No.
2021ZD0301900, National Natural Science Foundation of China (NSFC) under Grant No. 12374332, Project supported by Cultivation Project of Shanghai Research Center for Quantum Sciences Grant No. LZPY2024, Shanghai Science and Technology Innovation Action Plan Grant No. 24LZ1400800, and Jiaoda 2030 program WH510363001-1.
\bibliographystyle{unsrt}


\bibliography{ref}

\clearpage
\begin{widetext}
\appendix

\begin{center}
\textbf{Supplemental Materials}
\end{center}

\section{I. The effects of cavity loss, multi-mode, and finite temperature on the results}


\textbf{Exploring the Impact of Cavity Loss:} In practical scenarios, cavities are seldom perfect, inevitably leading to cavity losses. Here, we delve into the ramifications of cavity loss, denoted as $\kappa$, on our outcomes. Interaction with the environment via $\kappa$ results in the broadening of atomic spectra, represented by $\Delta \omega$. To maintain stability in the ground state of the light-matter interacting system $\hat H=\hat H_{\rm matt}+\hat H_{\rm photon}$, it's imperative that $\Delta \omega << E_{12}$, where $E_{12}$ signifies the energy difference between the ground state and the first excited state of the coupled system.

In the main text, we examine two scenarios: the weak coupling case and the strong coupling case, contingent upon the relative strengths of the light-matter coupling parameter $g$ and the cavity frequency $\omega_c$. Introducing cavity loss introduces three distinct energy scales: $g$, $\omega_c$, and $\kappa$, enriching the underlying physics. For a well-fabricated cavity (characterized by a high quality factor $Q$) with clearly separated modes, ensuring $\omega_c \gg \Delta \omega$ is generally feasible. Consequently, only three potential situations worth consideration here. i) $g\gg \omega_c \gg \Delta \omega$: In this instance, strong light-matter coupling predominantly dictates the physics, with cavity loss playing a negligible role. ii) $\omega_c \gg g\gg \Delta \omega$: Here, the light-matter coupling induces a spectral shift that remains prominent despite cavity losses. iii) $\omega_c \gg \Delta \omega \gg g$: In this scenario, cavity loss surpasses light-matter coupling, and the cavity Lamb shift becomes obscured by spectral broadening over a timescale determined by $\kappa^{-1}$.

\textbf{Further understand the influence of cavity loss from the Jaynes-Cummings model:} We can further understand the statement of the last paragraph through the well-known Jaynes-Cummings model. 
Under the rotating wave and single mode approximation [S1], the coupled qubit-cavity system can be effectively described by the Jaynes-Cummings Hamiltonian:
\begin{eqnarray}
\hat H=\hbar \omega_{ge}\hat \sigma_{ee}+\hbar \omega_r \hat a^\dagger \hat a+\hbar g_{ge}\left(\hat \sigma_{ge}^\dagger \hat a+\hat a^\dagger \hat \sigma_{ge}\right)
\end{eqnarray}
where $\omega_{ge}$ represents the transition frequency between the qubit's ground state $\lvert g \rangle$ and excited state $\lvert e \rangle$, and $\hat{\sigma}{ij} = \lvert i \rangle \langle j \rvert$ are the corresponding transition operators. $\omega_r$ is the resonant frequency of the cavity, with $\hat{a}^\dagger$ and $\hat{a}$ being the photon creation and annihilation operators acting on the photon number states $\lvert n \rangle$. The parameter $g{ge}$ characterizes the coupling strength between the qubit and the cavity.

We will now consider two limiting cases. In the resonant limit, where $\Delta = \omega_{ge} - \omega_r \rightarrow 0$, a single quantum of energy can be coherently exchanged between the qubit and the cavity. The system in this limit is described by the joint dressed states $\lvert n_{\pm} \rangle = \left( \lvert g, n \rangle \pm \lvert e, n-1 \rangle \right)/\sqrt{2}$.

A qualitatively different regime occurs when the qubit and cavity are far detuned from each other, specifically, when $|\Delta| \gg g_{ge}$. In this dispersive limit, the two systems do not exchange energy resonantly; instead, their interaction manifests as a spectral shift. The qubit transition frequency $\omega_{ge}$
is renormalized by the dispersive interaction, resulting in a detuning-dependent ac Stark shift [S2]:
\begin{eqnarray}
\delta_S \sim 2n \frac{g_{ge}^2}{\Delta}
\end{eqnarray}
for a cavity field populated with $n=\langle \hat a^\dagger \hat a \rangle$, as well as a Lamb shift
\begin{eqnarray}
\delta_L \sim \frac{g_{ge}^2}{\Delta}
\end{eqnarray}
due to the interaction with vacuum fluctuations 
Notably, the Lamb shift is insensitive to cavity loss because the photon number in the cavity is already zero.

\textbf{Multi-mode effect on the spectral shift:}
Here, we investigate the multi-mode effect on our result. We generalize the single-mode model in our main text to multi-mode case. The cavity QED single-particle Hamiltonian in the Coulomb gauge reads
\begin{equation}
\hat H_C= \frac{\left(\hat{\bold p}-q\hat{\bold A}\right)^2}{2m}+V(\bold r)+\sum_k\hbar \omega_k\hat a_k^\dagger \hat a_k
\end{equation}
where $V(\bold r)$ is an external single-particle potential. $\hat a$ denotes the annihilation operator of the bare photonic modes of frequency $\omega_c$, and the vector potential operator can be expressed as $\hat{\bold A}=\sum_k A_0\left(\boldsymbol\varepsilon \hat a_k +\boldsymbol \varepsilon^* \hat a_k^\dagger\right)$; here $\boldsymbol \varepsilon$ stands for the polarization of the cavity photonic modes. In this paper, we will consider the chiral cavity where the polarization is described $\boldsymbol\varepsilon=\frac{1}{\sqrt{2}}\left(\bold e_x+i\bold e_y\right)$. Expand the above Hamiltonian, and we obtain the following Hamiltonian
\begin{eqnarray}\label{hamilton1}
\hat H_C&=&\frac{\hat{\bold p}^2}{2m}+V(\bold r) - \sum_k\frac{qA_0}{m} \hat{\bold p}\cdot\left(\boldsymbol\varepsilon\hat a_k+\boldsymbol\varepsilon^*\hat a_k^\dagger\right)+\sum_k\hbar \omega_k\hat a_k^\dagger \hat a_k +\sum_{k_1,k_2} \frac{(qA_0)^2}{2m}\left(\hat a_{k_1}^\dagger\hat a_{k_2}+\hat a_{k_1}\hat a_{k_2}^\dagger\right)
\end{eqnarray}
Note that the last term stands for the mode mixing (or mode interference) effect, which is physically induced by cavity light-matter coupling.

To asymptotically decouple light and matter degrees of freedom, we propose to use a unitary transformation including multi-mode situation. 
\begin{equation}
\hat U=\Pi_k\exp\left[-i \frac{\xi_k}{\hbar} \hat{\bold p}\cdot \hat{\boldsymbol\pi_k}\right],
\end{equation}
where $\hat{\boldsymbol \pi}_k=i\left(\boldsymbol \varepsilon^* \hat a_k^\dagger-\boldsymbol \varepsilon \hat a_k\right)$ which has the meaning of photonic momentum operator. $\xi_k$ is to be determined so that the transformed Hamiltonian contains no term of linearly light-matter coupled term. 

One can write down the components of $\hat{\boldsymbol \pi_k}$ either in the presentation of Cartesian or polar coordinates. 
In Cartesian coordinates
\begin{eqnarray}
\hat{\pi}_{kx}=\frac{i}{\sqrt{2}}\left(\hat a_k^\dagger-\hat a_k\right);~~~\hat{\pi}_{ky}=\frac{1}{\sqrt{2}}\left(\hat a_k^\dagger+\hat a_k\right)
\end{eqnarray}
while in polar coordinates
\begin{eqnarray}
\hat{\pi}_{kr}&=&\hat{\boldsymbol \pi}_k\cdot \bold{e_r}=\frac{i}{\sqrt{2}}\left(e^{-i\theta}\hat a_k^\dagger-e^{i\theta}\hat a_k\right);\\
\hat{\pi}_{k\theta}&=&\hat{\boldsymbol \pi}_k\cdot \bold{e_\theta}=\frac{1}{\sqrt{2}}\left(e^{-i\theta}\hat a_k^\dagger+e^{i\theta}\hat a_k\right)
\end{eqnarray}
To calculate the following transformation
\begin{equation}
\hat H_{U}=\hat U^\dagger \hat H_C \hat U=\exp\left[\sum_k i \frac{\xi_k}{\hbar} \hat{\bold p}\cdot \hat{\boldsymbol\pi}_k\right] \hat H_C \exp\left[-\sum_k i \frac{\xi_k}{\hbar} \hat{\bold p}\cdot \hat{\boldsymbol\pi}_k\right],
\end{equation}
we often use the following identity (so-called Hadamard's lemma)
\begin{eqnarray}
e^{\hat A}\hat B e^{-\hat A}=\hat B+[\hat A, \hat B]+\frac{1}{2!}[\hat A, [\hat A, \hat B]]+\frac{1}{3!}[\hat A, [\hat A, [\hat A, \hat B]]]+\dots
\end{eqnarray}
Three transformations are of particular importance. 

i) The first transformation term
\begin{eqnarray}
\begin{aligned}
\hat U^\dagger \sum_k \hat{\bold p}\cdot\left(\boldsymbol\varepsilon^*\hat a_k^\dagger+\boldsymbol\varepsilon\hat a_k\right)\hat U
=&\sum_k\hat{\bold p}\cdot\left(\boldsymbol\varepsilon\hat a_k+\boldsymbol\varepsilon^*\hat a_k^\dagger\right)+\sum_{k_1,k}[\frac{i\xi_{k_1}}{\hbar} \hat{\bold p}\cdot \hat{\boldsymbol\pi_{k_1}}, \hat{\bold p}\cdot\left(\boldsymbol\varepsilon\hat a_k+\boldsymbol\varepsilon^*\hat a_k^\dagger\right)]+\dots\\
=&\sum_k\hat{\bold p}\cdot\left(\boldsymbol\varepsilon\hat a_k+\boldsymbol\varepsilon^*\hat a_k^\dagger\right)-\sum_{k,k_1}\frac{\xi_{k_1}}{\hbar}[ \hat{\bold p}\cdot \left(\boldsymbol \varepsilon^* \hat a_{k_1}^\dagger-\boldsymbol \varepsilon \hat a_{k_1}\right), \hat{\bold p}\cdot\left(\boldsymbol\varepsilon\hat a_k+\boldsymbol\varepsilon^*\hat a_k^\dagger\right)]+\dots\\
=&\sum_k\hat{\bold p}\cdot\left(\boldsymbol\varepsilon\hat a_k+\boldsymbol\varepsilon^*\hat a_k^\dagger\right)+2\sum_k\frac{\xi_k}{\hbar}\left(\hat{\bold p}\cdot \boldsymbol \varepsilon^*\right)\left(\hat{\bold p}\cdot \boldsymbol \varepsilon \right)
\end{aligned}
\end{eqnarray}

ii) The second transformation is
\begin{eqnarray}
\begin{aligned}
\hat U^\dagger\left(\sum_k \hat a_k^\dagger \hat a_k\right) \hat U=&\sum_k \hat a_k^\dagger \hat a_k
+{ [\sum_{k_1}\frac{i\xi_{k_1}}{\hbar} \hat{\bold p}\cdot \hat{\boldsymbol\pi}_{k1},\sum_k \hat a_k^\dagger\hat a_k]}
+{ \frac{1}{2!}[\sum_{k1}\frac{i\xi_{k1}}{\hbar} \hat{\bold p}\cdot \hat{\boldsymbol\pi_{k1}}, \sum_{k2}[\frac{i\xi_{k2}}{\hbar} \hat{\bold p}\cdot \hat{\boldsymbol\pi_{k2}},\sum_k\hat a_k^\dagger\hat a_k]]}+\dots\\
=&\sum_k\hat a_k^\dagger \hat a_k
+{ \sum_k \frac{\xi_k}{\hbar} \hat{\bold p}\cdot\left(\boldsymbol \varepsilon^*\hat a_k^\dagger+\boldsymbol \varepsilon \hat a_k\right)}
+ { \sum_k\frac{\xi_k^2}{\hbar^2}\left(\hat{\bold p}\cdot \boldsymbol \varepsilon^*\right)\left(\hat{\bold p}\cdot \boldsymbol \varepsilon \right)}
\end{aligned}
\end{eqnarray}

iii) The third transformation
\begin{eqnarray}
\begin{aligned}
\hat U^\dagger \hat{r}_i\hat U=&\hat{r}_i+[\sum_k\frac{i\xi_k}{\hbar} \hat{\bold p}\cdot \hat{\boldsymbol\pi}_k, \hat{r}_i]+\frac{1}{2!}[\sum_{k_1}\frac{i\xi_{k_1}}{\hbar} \hat{\bold p}\cdot \hat{\boldsymbol\pi}_{k_1}, \sum_{k_2}[\frac{i\xi_{k_2}}{\hbar} \hat{\bold p}\cdot \hat{\boldsymbol\pi}_{k_2}, \hat{r}_i]]+\dots\\
=&\hat{r}_i+\sum_k\xi_k \hat\pi_{ki}+\frac{1}{2!}\sum_{k_1,k_2}[\frac{i\xi_{k_1}}{\hbar} \hat{\bold p}\cdot \hat{\boldsymbol\pi}_{k_1}, \xi_{k_2} \hat\pi_{k_2i}]+\dots,
\end{aligned}
\end{eqnarray}
where $\hat \pi_{kx}=\frac{i}{\sqrt{2}}\left(\hat a_k^\dagger-\hat a_k\right)$, $\hat\pi_{ky}=\frac{1}{\sqrt{2}}\left(\hat a_k^\dagger+\hat a_k\right)$, and therefore $[\hat \pi_{kx},\hat\pi_{ky}]=-i$. Further we can obtain
\begin{eqnarray}
\begin{aligned}
\hat U^\dagger \hat{r}_i\hat U
=&\hat{r}_i+\sum_{k}\xi_k \hat\pi_{ki}+\frac{1}{2}[\sum_{k_1}\frac{i\xi_{k_1}}{\hbar} \hat{\bold p}\cdot \hat{\boldsymbol\pi}_{k_1}, \sum_{k_2}\xi_{k_2} \hat\pi_{k_2i}]+\dots\\
=&\hat{r}_i+\sum_{k}\xi_k \hat\pi_{ki}+\sum_k\frac{\xi_k^2}{2\hbar}p_j \epsilon_{ji},
\end{aligned}
\end{eqnarray}
where $\epsilon_{xy}=-\epsilon_{yx}=1$ and $\epsilon_{xx}=\epsilon_{yy}=0$. Compactly, one can write the above formula as
\begin{eqnarray}
\hat U^\dagger \hat{\bold r} \hat U
=\hat{\bold r}+\sum_k \xi_k \hat{\boldsymbol \pi}_k+\sum_k \frac{\xi_k^2}{2\hbar }\hat{\bold p}\times \bold{e_z}
\end{eqnarray}

iii) The fourth transformation
\begin{eqnarray}
\begin{aligned}
&\sum_{k_1,k_2}\hat U^\dagger\left(\hat a_{k_1}^\dagger \hat a_{k_2}+\hat a_{k_1} \hat a_{k_2}^\dagger\right) \hat U=\sum_{k_1,k_2}\hat U^\dagger\left(\hat a_{k_1}^\dagger \hat a_{k_2}+\hat a_{k_2}^\dagger \hat a_{k_1}+\delta_{k_1,k_2}\right) \hat U\\
&=N +\sum_{k_1,k_2} \left(\hat a_{k_1}^\dagger \hat a_{k_2}+\hat a_{k_2}^\dagger \hat a_{k_1}\right)
+{ [\sum_{k_1}\frac{i\xi_{k_1}}{\hbar} \hat{\bold p}\cdot \hat{\boldsymbol\pi}_{k1},\sum_{k_2,k_3} \left(\hat a_{k_2}^\dagger \hat a_{k_3}+\hat a_{k_3}^\dagger \hat a_{k_2}\right)]}\\
&
+{ \frac{1}{2!}[\sum_{k1}\frac{i\xi_{k1}}{\hbar} \hat{\bold p}\cdot \hat{\boldsymbol\pi}_{k1}, \sum_{k2}[\frac{i\xi_{k2}}{\hbar} \hat{\bold p}\cdot \hat{\boldsymbol\pi}_{k2},\sum_{k_3,k_4} \left(\hat a_{k_3}^\dagger \hat a_{k_4}+\hat a_{k_4}^\dagger \hat a_{k_3}\right)]]}+\dots\\
&=N+\sum_{k_1,k_2} \left(\hat a_{k_1}^\dagger \hat a_{k_2}+\hat a_{k_2}^\dagger \hat a_{k_1}\right)
+\sum_{k_1,k_2,k_3} \frac{i\xi_{k_1}}{\hbar} \hat{\bold p}\cdot { [ \hat{\boldsymbol\pi}_{k1}, \left(\hat a_{k_2}^\dagger \hat a_{k_3}+\hat a_{k_3}^\dagger \hat a_{k_2}\right)]}\\
&
+\sum_{k_1,k_2,k_3,k_4}\frac{i\xi_{k1}}{\hbar}\frac{i\xi_{k2}}{\hbar} \hat{\bold p}\cdot { \frac{1}{2!}[ \hat{\boldsymbol\pi}_{k1}, [\hat{\bold p}\cdot \hat{\boldsymbol\pi}_{k2}, \left(\hat a_{k_3}^\dagger \hat a_{k_4}+\hat a_{k_4}^\dagger \hat a_{k_3}\right)]]}+\dots\\
\end{aligned}
\end{eqnarray}
In the above expression, we assumed there are $N=\sum_k$ modes confined in the cavity.
The key equation one needs to calculate is 
\begin{eqnarray}
\begin{aligned}
[\hat{\boldsymbol\pi}_{k1}, \left(\hat a_{k_2}^\dagger \hat a_{k_3}+\hat a_{k_3}^\dagger \hat a_{k_2}\right)]&=[i\left(\boldsymbol \varepsilon^* \hat a_{k_1}^\dagger-\boldsymbol \varepsilon \hat a_{k_1}\right),  \left(\hat a_{k_2}^\dagger \hat a_{k_3}+\hat a_{k_3}^\dagger \hat a_{k_2}\right)]\\
&=i\boldsymbol\varepsilon^*[  \hat a_{k_1}^\dagger,  \left(\hat a_{k_2}^\dagger \hat a_{k_3}+\hat a_{k_3}^\dagger \hat a_{k_2}\right)]-i\boldsymbol \varepsilon [\hat a_{k_1}, \left(\hat a_{k_2}^\dagger \hat a_{k_3}+\hat a_{k_3}^\dagger \hat a_{k_2}\right)]\\
&=i\boldsymbol\varepsilon^*\left[\hat a_{k_2}^\dagger(-\delta_{k_1,k_3})+ \hat a_{k_3}^\dagger(-\delta_{k_1,k_2})\right]-i\boldsymbol \varepsilon  \left[\hat a_{k_3} \delta_{k_1,k_2}+\hat a_{k_2}\delta_{k_1,k_3}\right]
\end{aligned}
\end{eqnarray}
Therefore, the second and the third term in the fourth transformation read
\begin{eqnarray}
\begin{aligned}
&\sum_{k_1,k_2,k_3} \frac{i\xi_{k_1}}{\hbar} \hat{\bold p}\cdot {\left(i\boldsymbol \varepsilon^*\left[\hat a_{k_2}^\dagger(-\delta_{k_1,k_3})+ \hat a_{k_3}^\dagger(-\delta_{k_1,k_2})\right]-i\boldsymbol \varepsilon  \left[\hat a_{k_3} \delta_{k_1,k_2}+\hat a_{k_2}\delta_{k_1,k_3}\right] \right)}\\
&=\sum_{k_1,k_2} \frac{2\xi_{k_1}}{\hbar} \hat{\bold p}\cdot { \left(\boldsymbol\varepsilon^*\hat a_{k_2}^\dagger+\boldsymbol \varepsilon  \hat a_{k_2} \right)}=\Gamma \sum_k \hat{\bold p}\cdot \left(\boldsymbol\varepsilon^*\hat a_{k}^\dagger+\boldsymbol \varepsilon  \hat a_{k} \right),
\end{aligned}
\end{eqnarray}
where $\Gamma=\sum_k\frac{2\xi_{k}}{\hbar}$.
\begin{eqnarray}
\begin{aligned}
&\frac{1}{2}\sum_{k_1,k_2} [\left(\frac{i\xi_{k_1}}{\hbar} \hat{\bold p}\cdot \boldsymbol{\hat \pi}_{k_1}\right), \Gamma \sum_k \hat{\bold p}\cdot \left(\boldsymbol\varepsilon^*\hat a_{k_2}^\dagger+\boldsymbol \varepsilon  \hat a_{k_2} \right)]\\
&=\sum_k\frac{\Gamma \xi_k}{\hbar} (\hat{\bold p}\cdot \boldsymbol\varepsilon^*)(\hat{\bold p}\cdot \boldsymbol\varepsilon)=\frac{\Gamma^2}{2}(\hat{\bold p}\cdot \boldsymbol\varepsilon^*)(\hat{\bold p}\cdot \boldsymbol\varepsilon)
\end{aligned}
\end{eqnarray}

Therefore, the transformed Hamiltonian turns out to be
\begin{eqnarray}
\begin{aligned}
H_U=&\frac{\hat{\bold p}^2}{2m}+V(\hat{\bold r}+\sum_k \xi_k \hat{\boldsymbol \pi}_k+\sum_k \frac{\xi_k^2}{2\hbar }\hat{\bold p}\times \bold{e_z}) +\sum_k\hbar\omega_k \hat a_k^\dagger \hat a_k+\sum_k {\xi_k \omega_k} \hat{\bold p}\cdot\left(\boldsymbol \varepsilon^*\hat a_k^\dagger+\boldsymbol \varepsilon \hat a_k\right)\\
&-\sum_k\left(\frac{qA_0}{m}-\frac{(qA_0)^2}{2m}\Gamma\right)\hat{\bold p} \cdot\left(\boldsymbol\varepsilon^*\hat a_k^\dagger+\boldsymbol\varepsilon\hat a_k\right)+\left(\frac{\Gamma^2}{2}-\frac{qA_0\Gamma}{m}+\sum_k\frac{\omega_k \xi_k^2}{\hbar}\right)\left(\hat{\bold p}\cdot \boldsymbol \varepsilon^*\right)\left(\hat{\bold p}\cdot \boldsymbol \varepsilon \right)\\
&+ \sum_{k_1,k_2} \left(\hat a_{k_1}^\dagger \hat a_{k_2}+\hat a_{k_2}^\dagger \hat a_{k_1}\right)\\
=&\frac{\hat{\bold p}^2}{2m}+V(\hat{\bold r}+\sum_k \xi_k \hat{\boldsymbol \pi}_k+\sum_k \frac{\xi_k^2}{2\hbar }\hat{\bold p}\times \bold{e_z}) +\sum_k\hbar\omega_k \hat a_k^\dagger \hat a_k+\sum_k {\xi_k \omega_k} \hat{\bold p}\cdot\left(\boldsymbol \varepsilon^*\hat a_k^\dagger+\boldsymbol \varepsilon \hat a_k\right)\\
&-\sum_k\left(\frac{qA_0}{m}-\frac{(qA_0)^2}{2m}\Gamma\right)\hat{\bold p} \cdot\left(\boldsymbol\varepsilon^*\hat a_k^\dagger+\boldsymbol\varepsilon\hat a_k\right)+\left(m\frac{\Gamma^2}{2}-{qA_0\Gamma}+\sum_k\frac{m\omega_k \xi_k^2}{\hbar}\right)\left(\frac{\hat{\bold p}^2}{2m} \right)\\
&+ \sum_{k_1,k_2} \left(\hat a_{k_1}^\dagger \hat a_{k_2}+\hat a_{k_2}^\dagger \hat a_{k_1}\right)\\
=&\frac{\hat{\bold p}^2}{2m_{d}}+V(\hat{\bold r}+\xi \hat{\boldsymbol\pi }+\frac{\xi^2}{2\hbar }\hat{\bold p}\times \bold{e_z}) +\sum_k\hbar\omega_k \hat a_k^\dagger \hat a_k+\sum_{k_1,k_2} \left(\hat a_{k_1}^\dagger \hat a_{k_2}+\hat a_{k_2}^\dagger \hat a_{k_1}\right)\\
&+\sum_k \left[\xi_k \omega_k - \left(\frac{qA_0}{m}-\frac{(qA_0)^2}{2m}\Gamma\right)\right] \hat{\bold p}\cdot\left(\boldsymbol \varepsilon^*\hat a_k^\dagger+\boldsymbol \varepsilon \hat a_k\right)
\end{aligned}
\end{eqnarray}
where $\frac{1}{m_d}=\frac{1}{m}\left(\frac{m\Gamma^2}{2}-2qA_0 \Gamma+\sum_k\frac{m \omega_k \xi_k^2}{\hbar}\right)$. It is important to note that, to eliminate the linear term, we require 
\begin{equation}\label{admultimode}
 \xi_k\omega_k=\frac{qA_0}{m}-\frac{q^2A_0^2}{2m}\Gamma
\end{equation}
and the transformed Hamiltonian becomes
\begin{eqnarray}\label{eqtran}
\hat H_U=\frac{\hat{\bold p}^2}{2m_{d}}+V\left(\hat{\bold r}+\sum_k \xi_k \hat{\boldsymbol \pi}_k+\sum_k \frac{\xi_k^2}{2\hbar }\hat{\bold p}\times \bold{e_z}\right)+\sum_k \hbar \omega_k \hat a_k^\dagger\hat a_k+ \sum_{k_1,k_2} \left(\hat a_{k_1}^\dagger \hat a_{k_2}+\hat a_{k_2}^\dagger \hat a_{k_1}\right).
\end{eqnarray}
The transformed Hamiltonian embodies all the cavity light-matter coupling into the shifted potential, which is an essential starting point of our paper. Now one key question comes: Will the equation always \eqref{admultimode} has a solution? Not necessarily, but in some cases it has. For example, one could consider a two-mode case, where one has
\begin{eqnarray}
\xi_1\omega_1&=&\frac{qA_0}{m}-\frac{q^2A_0^2}{2m}\left(\frac{2\xi_1}{\hbar}+\frac{2\xi_1}{\hbar}\right)\\
\xi_2\omega_2&=&\frac{qA_0}{m}-\frac{q^2A_0^2}{2m}\left(\frac{2\xi_1}{\hbar}+\frac{2\xi_1}{\hbar}\right)
\end{eqnarray}
The solutions of the above set of equation are
\begin{eqnarray}
\xi_1&=&\frac{\omega_2\left(\frac{qA_0}{m}\right)}{\omega_1\omega_2+\left(\frac{q^2A_0^2}{2m}\right)(\omega_1+\omega_2)}\\
\xi_2&=&\frac{\omega_1\left(\frac{qA_0}{m}\right)}{\omega_1\omega_2+\left(\frac{q^2A_0^2}{2m}\right)(\omega_1+\omega_2)}
\end{eqnarray}
Extending to the $N$-mode case, the general solution turns out to be
{ \begin{eqnarray}\label{multieqtran}
\xi_i=\frac{\prod^N_{k\neq i}\omega_k\left(\frac{qA_0}{m}\right)}{\prod_{k=1}^N\omega_k+\left(\frac{q^2A_0^2}{2m}\right)\left[\sum_{j=1}^N \prod_{k\neq j}\omega_k\right]}
\end{eqnarray}
}
This result indicates that no matter how many modes we consider, there are always solutions for $\xi$. It is a remarkable result. Before moving on, let us dwell a bit to check dimensionality of above formula: $[qA_0]= [\text{momentum}]$ and $[\xi]=[\text{distance}]$ indicate that everything in \eqref{multieqtran} are correct in unit. 

Another fact that worthy noting is the expression of the parameter $\xi$: In the weak coupling limit, $A_0\rightarrow 0$ ($A_0$ is a dimensional quantity. Here it actually means $\frac{(qA_0)^2}{m}<<\hbar \omega_c$), $\xi\rightarrow \frac{qA_0}{m\omega_i}$; whereas in the strong coupling limit, $A_0\rightarrow \infty$ (actually means $\frac{(qA_0)^2}{m}>>\hbar \omega_k$), and $\xi\rightarrow \frac{2\prod^N_{k\neq i}\omega_k }{qA_0 \left(\sum_{j=1}^N \prod_{k\neq j}\omega_k\right)}$. In either case, $\xi$ is a comparably small length scale, which indicates that perturbation is legitimately allowed when the length scale we care about is larger than $\xi$. 

Additional remark: One often define a coupling constant $g\equiv\sqrt{\frac{(qA_0)^2}{m}/\hbar \omega_c}$, because it is dimensionless and linearly proportional to the strength $qA_0$. it is often called strong coupling in cavity QED context, if $10^{-1}\leqslant g\leqslant 1$; it is called deep strong coupling regimes for $g\geqslant 1$ [S3]. In terms of the dimensionless coupling constant $g$, the parameter (with dimension [length]) 
\begin{eqnarray}
\xi=\frac{g}{1+g^2}\sqrt{\frac{\hbar}{m\omega_c}}
\end{eqnarray}






\textbf{Finite temperature effect:}
Another thing that we need to discuss is the photon population due to thermal fluctuations. In this realistic experiment [S4], the system is in a temperature $T_r=90 {\rm mK}$. Assuming a cavity resonant frequency $\omega_r/2\pi=6.5 \times 10^9  {\rm Hz}$, we can calculate the averaged populated photon number
\begin{eqnarray}
    \bar n=\frac{k_B T_r}{\hbar \omega_r}=\frac{\frac{k_B T_r}{k_B (90 ~{\rm m K})}}{\frac{\hbar \omega_r}{\hbar 6.5 {\rm GHz}}}\times \frac{k_B (90 {\rm mK})}{\hbar (6.5 {\rm GHz})}= \frac{T_r}{90 {\rm mK}} \times \frac{6.5 ~{\rm GHz}}{\omega_r}\times 0.03
\end{eqnarray}
Therefore, the thermal contribution is small compared to the vacuum fluctuations contribution.

\section{II. Characterization of chiral cavity and its experimental implementation}

\textbf{Cavity enhanced quantum fluctuations:} The amplification of quantum fluctuations in the cavity can be described by the  Hamiltonian of quantum electrodynamics for a single mode within a cavity of volume $V$:
$$
\hat H_{\text{cavity}}=\frac{\epsilon_0 V}{2}(E^2+c^2 B^2)=\frac{\epsilon_0 V}{2}({\dot A}^2+c^2 |\bold k\times\bold A|^2)
$$
This Hamiltonian can be directly mapped onto the Hamiltonian of a harmonic oscillator 
$$\hat H_{\text{HO}}=\frac{m}{2} (\dot x^2+\Omega^2x^2)
$$
by the substitution $A\mapsto x $, $\epsilon_0 V\mapsto m$, and $ck\equiv \omega_c\mapsto \Omega$.
The ground state of a harmonic oscillator exhibits quantum fluctuations, quantified by the root-mean-square standard deviation of position: $\Delta x = \left(\frac{\hbar}{2\Omega m}\right)^{\frac{1}{2}}$. This shows that the vacuum fluctuations of the vector potential are given by $\Delta A=(\hbar/2\omega_c\epsilon_0 V)^{\frac{1}{2}}$. Clearly, reducing the cavity volume can amplify quantum fluctuations significantly. In this section, we will detail the necessary conditions for creating a chiral cavity capable of inducing the effects outlined in the main text.

\textbf{Chiral cavity model and experimental implementation:}
Experimentally, there are different ways to fabricate chiral cavities. And in fact, many papers interchangeingly use chiral and ``gyrotropic", which we believe is not accurate enough. In our definition, chiral cavity represents a cavity that breaks time-reversal symmetry, whereas a gyrotropic cavity only breaks parity symmetry. As we have demonstrated in the main text, we propose to use a Faraday rotator material deposited on top of a metal to make a chiral cavity. The metallic substrate is required for reducing photon loss (namely, the Q factor) of the cavity. 
The cavity modal volume determines the coupling strength between light and atom. And the proposed 
cavity size is quite conservative in the main text, with an effective mode volume of approximately $(1~\mathrm{\mu m})^3$. Notably, experiments have demonstrated the possibility of building much smaller cavities with an effective mode volume on the order of $V_{\rm eff}\sim 1\times 10^{-5} \times(\lambda/2\sqrt{\epsilon})^{3}$, where $\lambda$ represents the wavelength of the confined photonic mode [S5]. These cavities have been utilized in experiments to investigate Landau polaritons [S6]. Due to their proven experimental feasibility, these effective parameters were also widely adopted in various theoretical papers [S7,S8].
Such chiral cavities have already been realized in experiments [S9], and a detailed theoretical analysis of these cavities can be found, for example, in [S10].

Finally, we remark that our proposed setup does not require a high $Q$ factor, as the mechanism in our proposal does not depend on resonant light-matter coupling. This significant distinction (compared to quantum optics setups) enhances the feasibility of utilizing a cavity to control condensed matter systems. Recently, a plethora of experimental works has emerged to explore cavity many-body systems. Among these experiments, two notable works stand out. One demonstrates that a vacuum cavity can break down the topological protection of quantum hall systems [S11], while the second illustrates that a cavity can be employed to control the metal-to-insulator transition of transition metal dichalcogenide material $\rm Ta S_2$ [S12]. All in all, the existing evidence supports the experimental feasibility of our proposal.

\textbf{Defining the chirality operators:}
In this part, we review how to define the chirality operator in terms of photonic creation and annihilation operators, following the elegant work [S13]. 
A chiral cavity is made of ferromagnetic metals which is described by 
the following effective Hamiltonian:
\begin{equation}
H_{E M}=H_{E M}^{(0)}+\frac{1}{2 c^2} \int d^3 \mathbf{r} \sigma_H(z)[\dot{\mathbf{A}} \times \mathbf{A}],
\end{equation}
where the ferromagnetic metals give rise to the second Hall effect term. To quantized the Hamiltonian, we start from quantizing vector potential:
$$
\hat{A}_{x, y}(z)=\sqrt{\frac{\hbar c^2}{2 V \Omega_0}}\left(\hat{a}_{x, y}+\hat{a}_{x, y}^{\dagger}\right) \phi(z)=\sqrt{\frac{\hbar c^2}{V \Omega_0}} \hat{q}_{x, y} \phi(z),
$$
where $V=S D$ is the mode volume, $S$ is the cavity area, and $\phi(z)$ is the normalized mode profile. Operators $\hat{a}_{x, y}$ are the conventional bosonic annihilation operators. The position operator is defined as 
$\hat q_i=\frac{1}{\sqrt{2}}\left(\hat a_i+\hat a_i^\dagger\right)$ where as the momentum operator is defined as 
$\hat \pi_i=-\frac{i}{\sqrt{2}}\left(\hat a_i-\hat a_i^\dagger\right)$ so that $\hat q_i$ and $\hat\pi_i$ is a pair of conjugate operators and they satisfy $[\hat q_i,\hat \pi_j]=i\delta_{ij}$.
\begin{equation}\label{chiraleigmod}
\begin{aligned}
H_{\mathrm{EM}} & =\hbar \Omega_0\left(\hat{a}_x^{\dagger} \hat{a}_x+\hat{a}_y^{\dagger} \hat{a}_y+1\right)+i \hbar \Delta\left(\hat{a}_x^{\dagger} \hat{a}_y-\hat{a}_y^{\dagger} a_x\right) \\
& =\frac{\hbar \Omega_0}{2} \hat{\boldsymbol{\pi}}^2+\frac{\hbar \Omega_0}{2} \mathbf{\hat q}^2+\hbar \Delta \hat{\boldsymbol{\pi}}\cdot\left(\boldsymbol{e}_z \times \mathbf{\hat q}\right),\\
&=\frac{\hbar \Omega_0}{2} \hat{\boldsymbol{\pi}}^2+\frac{\hbar \Omega_0}{2} \mathbf{\hat q}^2+\hbar \Delta \boldsymbol{e}_z \cdot \left(\mathbf{\hat q}\times\hat{\boldsymbol{\pi}}\right),
\end{aligned}
\end{equation}
Therefore, it is straightforward to define the chirality operator 
\begin{eqnarray}
\hat \chi=\hat {\bold q}\times {\hat{\boldsymbol \pi}}
\end{eqnarray}

As a side comment, Eq. \eqref{chiraleigmod} is the reason why the chiral mode is the eigen mode. From this equation, we see $\hat H_{EM}=\hbar\Omega_0\left(\hat a_x+i\lambda \hat a_y\right)\left(\hat a_x-i\lambda \hat a_y\right)$, where $\lambda=\Delta/\Omega_0$. The vector potential operator is $\hat{\vec A}=\hat a_x \vec e_x+\hat a_y \vec e_y+h.c.$. If we redefine $\hat b_x^\dagger=\frac{1}{\sqrt{2}}\left(\hat a_x^\dagger+i\lambda \hat a_y^\dagger\right)$ and $\hat b_x=\frac{1}{\sqrt{2}}\left(\hat a_x-i\lambda \hat a_y\right)$, and correspondingly $\hat b_y^\dagger=\frac{1}{\sqrt{2}}\left(\hat a_y^\dagger+i\lambda \hat a_x^\dagger\right)$.   In terms of $\hat a^\dagger$ and $\hat a$, the vector potential reads $\hat{\vec A}= A_0\left[a^\dagger \left(\vec e_x+i \vec e_y\right)+a \left(\vec e_x-i \vec e_y\right)\right]$.

\section{III. The second order correlation function of cavity photons at zero temperature}

To quantitatively characterize the quantum nature of the cavity fields, we follow the standard approach of calculating the second-order correlation function of photonic operators. By analyzing the photon statistics---whether sub-Poissonian or Poissonian---through the calculation of the second-order correlation function, we can determine whether the cavity field exhibits quantum or classical characteristics. In fact, there are quite a few works studying the statistical nature of cavity fields [S14-S18]. We need to trace out the matter degree of freedom to understand how atoms may influence statistic properties of cavity photons. Our calculations reveal that light-matter interactions cause the cavity photons to continue obeying a Poisson distribution. Therefore, taking into account the photon-atom interaction, the cavity photonic fields exhibit classical behavior. 

A coherent state for a photonic system is defined as
$\hat a |\alpha\rangle=\alpha |\alpha\rangle$
and correspondingly
$ \langle \alpha| \hat a^\dagger = \langle\alpha|\alpha^*$.
Expanding a coherent state in terms of photonic number states, one obtains
\begin{eqnarray}
    |\alpha\rangle=e^{-\frac{|\alpha|^2}{2}}\sum_{n=0}^{\infty} \frac{\alpha^n}{\sqrt{n!}}|n\rangle
\end{eqnarray}
For the single mode case, one can use the normalized second order correlation function to characterize nonclassicality of photonic field
\begin{eqnarray}
 g^{(2)}(\tau)=g^{(2)}(0)=\frac{\langle \hat a^\dagger \hat a^\dagger \hat a \hat a \rangle}{\langle \hat a^\dagger \hat a \rangle^2} = 1+ \frac{\langle (\Delta \hat n)^2 \rangle-\langle \hat n\rangle }{\langle \hat n \rangle^2}
\end{eqnarray}
Alternatively, one could define the so-called Q-parameter $Q=\langle n\rangle \left(g^{2}(0)-1\right)$. Through the second-order correlation function being $g^{(2)}$, one could find out the statistics of photons, namely, sub-Poissonian($g^{(2)}<1$), Poissonian ($g^{(2)}>1$), or super-Poissonian ($g^{(2)}>1$).

Given the transformed Hamiltonian,
\begin{eqnarray}
    \hat H=\frac{\hat p^2}{2 m_{\rm eff}}+V\left(r+\xi \hat{\boldsymbol \pi}+\frac{\xi^2}{2\hbar}\hat{\boldsymbol p}\times \boldsymbol{e_z}\right)+\hbar \omega_{\rm eff} \hat a^\dagger \hat a
\end{eqnarray}
In the strong coupling limit, within the condition that $\omega_{\rm eff}$ much larger than other energy scales in the problem, the ground state of the transformed Hamiltonian is simply a product state of the matter part and the photonic part
\begin{eqnarray}
|G\rangle_{\rm tran}=|\psi_{\rm matt}\rangle \otimes |0\rangle
\end{eqnarray}
which implies the real ground state of the coupled light matter system is 
\begin{eqnarray}
    |G\rangle =e^{i\frac{\xi}{\hbar} \hat{\boldsymbol p}\cdot \hat{\boldsymbol \pi}}|\psi_{\rm matt}\rangle \otimes |0\rangle. 
\end{eqnarray}
On the other hand, we know the vacuum displacement operator in quantum optics textbook, which is defined as
\begin{eqnarray}
    \hat D(\alpha)= \exp\left(\alpha \hat a^\dagger -\alpha^* \hat a\right)
\end{eqnarray}
With the definition of displacement operator, the coherent state can be written in terms of the vacuum state:
\begin{eqnarray}
    |\alpha\rangle =\hat D(\alpha)|0\rangle,
\end{eqnarray}
where, again, $|0\rangle$ denotes the vacuum state. The operator in our case turns out to be a ``super" displacement operator $\hat D(\hat \alpha)$ with $\hat \alpha=-\frac{\xi}{\hbar} \hat{\boldsymbol p}\cdot{\boldsymbol \epsilon}^*$. Notice that here $\hat \alpha$ is an operator here. For a special quantum states of the matter part, $\hat \alpha$ is replaced by its average in the matter state: $\hat \alpha \rightarrow \langle \psi_{\rm matt}| \hat \alpha |\psi_{\rm matt}\rangle$.
Therefore, given the definite quantum state of the matter (which is an atom in our case), the second order correlation reads
\begin{eqnarray}
g^{(2)}(0)=\frac{\langle\psi_{\rm matt}|\otimes \langle 0| \hat a^\dagger \hat a^\dagger \hat a \hat a |0\rangle\otimes |\psi_{\rm matt}\rangle}{\langle\psi_{\rm matt}|\otimes \langle 0| \hat a^\dagger \hat a |0\rangle\otimes |\psi_{\rm matt}\rangle^2} = 1+ \frac{\langle (\Delta \hat n)^2 \rangle-\langle \hat n\rangle }{\langle \hat n \rangle^2}=1
\end{eqnarray}
This implies that one can devise a corresponding classical physical picture to capture the physics we discussed, provided we know the number of photons hidden in the ground state due to light-matter interaction. We will leave this issue for further study. However, we should point out that when cavity photons couple to a many-body system, light-matter entanglement emerges. Integrating out the matter degrees of freedom can produce squeezed light, which is best described by quantum theories [S19].

\begin{center}
\textbf{Supplemental References}
\end{center}

[S1] 
A. M. Fox, Quantum optics: an introduction, Vol. 15 (Oxford University Press, USA, 2006).

[S2]
M. Brune, P. Nussenzveig, F. Schmidt-Kaler, F. Bernardot, A. Maali, J. M. Raimond, and S. Haroche,
Phys. Rev. Lett. {\bf 72}, 3339 (1994)

[S3]
Kanta Masuki and Yuto Ashida,
Phys. Rev. B {\bf 107}, 195104 (2023)

[S4]
A. Wallraff, {\it et al.} Nature {\bf 431}, 162-167 (2004).

S[5]
G. Scalari,{\it et al.}  Journal of Applied Physics {\bf 113}, 136510 (2013)

[S6]
Curdin Maissen, {\it et al.}
Phys. Rev. B {\bf 90}, 205309 (2014)

[S7]
F. Schlawin, A. Cavalleri, and D. Jaksch, Phys. Rev. Lett. {\bf 122}, 133602 (2019).

[S8]
Xiao Wang, Enrico Ronca, and Michael A. Sentef, Phys. Rev. B {\bf 99}, 235156 (2019).

[S9]
H. H{\" u}bener {\it et al.}  Nature materials {\bf 20}, 438 (2021).

[S10]
L. Mauro, J. Fregoni, J. Feist, and R. Avriller, Phys. Rev. A {\bf 107}, L021501 (2023).

[S11]
F. Appugliese {\it et al.}, Science, {\bf 375}, 1030 (2022).

[S12]
Giacomo Jarc {\it et al.}, Nature {\bf 622}, 487-492 (2023).

[S13] I. V. Tokatly, D. R. Gulevich, and I. Iorsh, Phys. Rev. B {\bf 104}, L081408 (2021).

[S14]
Ivan Amelio, Lukas Korosec, Iacopo Carusotto, and Giacomo Mazza, Phys. Rev. B {\bf 104}, 235120 (2021).

[S15]
J. Phys. Chem. Lett. 14, 51, 11725-11734 (2023).

[S16]
V. Rokaj, S. I. Mistakidis, H. R. Sadeghpour, SciPost Phys. {\bf 14}, 167 (2023).

[S17]
Davis M. Welakuh, Michael Ruggenthaler, Mary-Leena M. Tchenkoue, Heiko Appel, and Angel Rubio,
Phys. Rev. Res. {\bf 3}, 033067 (2021).

[S18]
Martin Kiffner, Jonathan R. Coulthard, Frank Schlawin, Arzhang Ardavan, and Dieter Jaksch,
Phys. Rev. B 99, 085116 (2019).

[S19] Giacomo Passetti, Christian J. Eckhardt, Michael A. Sentef, and Dante M. Kennes, Phys. Rev. Lett. {\bf 131}, 023601 (2023).

\end{widetext}

\end{document}